\documentstyle[prb,aps,epsf,preprint]{revtex}
\tightenlines
\begin{document}
\title{Magnetic properties of correlated electrons}
\author{Romuald Lema\'{n}ski}
\address{Institute of Low Temperature and Structure Research, \\
Polish Academy of Sciences, Wroc\l aw, Poland}
\maketitle
\begin{abstract}
We analyze a system composed of itinerant electrons and localized spins with on-site
interactions representing the Hund's first rule. Properties of the system are studied
rigorously on the infinite square lattice but the configurational space is restricted
to the lowest-period phases only. Using exact expressions for the ground state energy
an evolution of the phase diagram with an external magnetic field is determined.
For a wide range of electron densities metamagnetic phase transitions are detected.
It is shown that jumps of magnetization of the electrons at the metamagnetic transitions
are much smaller than changes of magnetization of the localized spins.
\end{abstract}
\section{Introduction}
In this paper we consider a model of localized spins and moving electrons on the square lattice.
The spins do not interact between themselves directly, but only by means of the
electrons. The interaction is local (on-site), however, since the electrons move,
it spreads out over the whole system. According to the Hund's first rule we assume
the coupling between the localized spins and the spins of moving electrons to be ferromagnetic
and, for simplicity, we choose the Ising-type anisotropy of the interaction.

The model can be viewed as a simplified version of the Falicov-Kimball model with the
Ising on-site interaction we studied in our precedent papers \cite{RL}. Indeed, if in the later
model we impose exactly one localized {\emph f-electron} per site then the only relevant terms
that left are those given in the first line of the Hamiltonian (\ref{ham}).
However, now we include also Zeeman type terms describing action of the magnetic field
on both the localized spins and the electrons.

A type of magnetic order results from a competition between all the terms of the Hamiltonian.
The Hund's term and the magnetic field terms tend to align all the spins ferromagnetically.
On the other hand, the kinetic energy term gives a favour of the antiferromagnetic
order at half filling and close to that point.

The model Hamiltonian has the following form.
\begin{eqnarray}
\label{ham}
H= & t\sum\limits _{\left\langle i,j \right\rangle}\sum\limits _{\sigma=\uparrow ,\downarrow}
d^+_{i,\sigma} d_{j,\sigma}
-\mu\sum\limits _{i}(n^d_{i,\uparrow}+n^d_{i,\downarrow} )
 -J \sum\limits _{i}(n^d_{i,\uparrow}-n^d_{i,\downarrow} ) s^z_{i}  \nonumber \\
& -h \sum\limits _{i}(n^d_{i,\uparrow} -n^d_{i,\downarrow} )
-h \sum\limits _{i}s^z_{i} ,  \;
\end{eqnarray}
where $<i,j>$ means the nearest neighbor lattice sites $i$ and $j$,
$\sigma$ is a spin index, $d_{i,\sigma}$  ($d^+_{i,\sigma}$) is an annihilation
(creation) operator, and $n^d_{i,\sigma}$ is an occupation number of itinerant electrons.
The on-site interaction between localized spins and itinerant
electrons which reflects the Hund's rule force is represented by the Ising-type
coupling constant $J$. The other parameters are: the hopping amplitude $t$ (in the sequel we
measure all energies in units of t) and the chemical potential $\mu$.

In order to simulate various dopings of the \emph{d-electrons} we allow for any value of their
density, i.e. $0\leq \rho_d=\rho _{d\uparrow}+\rho _{d\downarrow}\leq 2$.
Consequently, at a given site the $d-electron$ occupancy $n_d=n_{d,\uparrow} + n_{d,\downarrow}$
is assumed to be equal to 0, 1 or 2.
So there are 2 states per site allowed for the spins ($\uparrow$ and $\downarrow$)
and 4 states per site allowed for the $d-electrons$
($n_d=0$; $n_{d,\uparrow}=1$ and $n_{d,\downarrow}=0$;
$n_{d,\uparrow}=0$ and $n_{d,\downarrow}=1$; $n_d=2$).

The first two terms of the Hamiltonian (\ref{ham}) describe kinetic energy of the electrons
in the ground canonical enesmble,
the third one - on-site Hund interaction that couples localized spins and
spins of itinerant electrons, and the last two terms represent energies
of spins and electrons in an external magnetic field.

The localized spins play a role of an external potential for the electrons.
This potential tunes its shape in such a way, that the total energy
of the system attains its minimum. So there is a feedback between the subsystems
of spins and electrons, and this is the feedback
that is responsible for long-range arrangements of the spins,

Our purpose here is to examine how ground-state arrangements of the
spins evolve when an external magnetic field $h$ is applied. We solve the problem
by constructing a restricted phase diagram of the model (\ref{ham})
within the configurational space, composed of all periodic phases
(and their mixtures), for which the number of sites per unit cell
is less or equal to 4. Then, for a chosen densities
of electrons, we calculate magnetization of the electrons as a function
of the magnetic field.
\section{Method of calculation}
Our trial set contains 12 periodic configurations of the spins. For each of them
we perform the Fourier transformation of the Hamiltonian (\ref{ham}) and determine
the electronic band structure of the conduction electrons. In other words, we solve
the eigenvalue problem and find the eigenvalues $E_{\nu \sigma}(k_x,k_y)$,
with branch index $\nu =1,2,...,N_0\leq 4$, spin index $\sigma $
and the Bloch wavevector $k=(k_x,k_y)$
(for more details see Refs. \cite{WatsonLemanski,LemanskiFreericksBanach}).
This requires us to diagonalize up to $4\times 4$ matrices and result
in analytical formulae for at most 4 different energy bands,
separately for spin-up and spin-down electrons.

In order to determine the ground-state energy we use a Brillouine
zone grid of momentum points (typicaly $100\times 100$) and sum eigenvalues of
each band stucture for each number of the conduction electrons.
Then we construct the ground canonical phase diagram by direct comparing
the Gibbs thermodynamical potentials of all phases from the trial set
and selecting the lowest one. Next, we translate the grand-canonical
diagram to the canonical phase diagram for arbitrary densitiy
$\rho_d=\rho _{d\uparrow}+\rho _{d\downarrow}$ of the $d-electrons$.
This procedure assures thermodynamical stability of all phases (both periodic
and their mixtures) present in the resulting canonical phase diagrams
\cite{GajekJedrzejewskiLemanski}.

In the current study we repeat the calculations for a discrete set of values
of the magnetic field $h$, starting from $0$ and up to $h_{max}=0.04$,
with the steps equal to $0.002$.
\section{Phase diagram}
The phase diagram (Fig. 1) has been constructed within the restricted space for $J=0.5$.
Apart from the ferromagnetic (F) and the simplest antiferromagnetic (AF) we have found
five other periodic phases in the diagram: two ferrimagnetic and three
antiferromagnetic. All the phases, but one ferrimagnetic, are stable already at $h=0$.

It is obvious that with an increase of the magnetic filed $h$ antiferromagnetic phases
become energetically less favorable than ferrimagnetic and ferrimagnetic less favorable than ferromagnetic.
This general rule is also observed in our phase diagram, however, details of the phase
transformations are not so evident. The simplest situation appears for electron densities close to
$0$ or $2$, where the F phase is stable already for $h=0$. The opposite limit is for $\rho _d\cong 1$,
where the AF phase persists up to relatively high values of $h$, and then transforms into the F phase
(the transition is not shown in the diagram).

More complex situation is observed for intermediate values of $\rho _d$, where either a ferrimagnetic
or an antiferromagnetic phase is a ground state for $h=0$. Then, even a tiny magnetic filed $h$
is able to invoke a metamagnetic transition to the F phase, or a series of the transitions:
first to a ferrimagnetic and then to the F phase (see Fig. 1,2).
\section{Other results and conclusions}
The present studies show that even a tiny magnetic field can cause a substantial change of
magnetic structure, especially for intermediate dopings. Although the range of densities
$\rho _d$, where phases other than F and AF are ground states shrinks with an increase of $h$,
some field induced arrangements (like one of the ferrimagnetic phases found in the present study)
can appear on the diagram. It suggests that metamagnetic phase transitions are perhaps common phenomena
in doped magnetic systems. Then, it would be interesting to verify experimentally the conjecture
by examination properties of relevant solid solutions in magnetic fields.

A reorientation of the spins at the metamagnetic phase transitions are accompanied by jumps of magnetization
of the electrons. However, the jumps are much smaller than the change of magnetization
of the spins, as it can be noticed in Fig. 3, where the magnetization versus magnetic
field for $J=0.5$ and two different electron desities: $\rho _d=0.3$, and $\rho _d=0.55$ (i.e. along the red
lines in Fig. 1) is shown.

It is also interesting that for some intervals of $h$ the magnetization of the electrons remains constant.
It means that in this intervals the density of states of the electrons has energy gaps.

Let us finally mention that the phase diagram restricted to the lowest period phases provides only
a crude information on the full diagram of the model, but we expect that this is the most essential
information. The conjecture is justified by the results found previously for the spinless
Falicov-Kimball model, where an increase in size of allowed unit cells does not produce
significant qualitative changes in the phase diagram \cite{WatsonLemanski,LemanskiFreericksBanach}.

\vspace{0.3cm}
\noindent
{\bf Acknowledgement}
  The work financed from the resources of the Polish State Committee for Scientific
  Research (KBN) in the period 2004-2006 as a scientific project No. 1 P03B 031 27.

\newpage

\begin{figure}
\caption{Phase diagram of the model for $J=0.5$. Only the discret values of the magnetic field
are taken into account: $h=0.002p$, where $p=0,1,2,...,20$. The horizontal straight line
segments show electron density ranges inside which particular phases are stable. Different
colours represent different phases: black - the ferromagnetic (F), red - the simplest
antiferromagnetic (AF). All other phases are displayed in Fig. 2 (their colours
are related to the colours of corresponding straight line segments). Blank gaps
between the stright line segments indicate those density intervals in which macroscopic
mixtures of two periodic phases have lower energy than any of them. The vertical red lines
indicate electron densities $\rho _d=0.3$ and $\rho _d=0.55$, for which we calculated
the magnetization of electrons as a function of the magnetic field (see Fig. 3).}
\label{fig1}
\end{figure}

\begin{figure}
\caption{Arrangements of localized spins in phases represented in the phase diagram displayed in Fig. 1
(the background colours identify the phases on the diagram).}
\label{fig2}
\end{figure}

\begin{figure}
\caption{Magnetization of the electrons versus magnetic field for $J=0.5$
and two different electron densities: $\rho _d=0.3$ (left panel) and $\rho _d=0.55$ (right panel).
Colours of the dots indicate stable arrangements of the localized spins
(see captions to Fig.1 and 2 for explanations). The lines are only guides to eyes.}
\label{fig3}
\end{figure}

\end{document}